\documentstyle[12pt]{article}
\newlength{\pubnumber} 

\catcode`\@=11
\@addtoreset{equation}{section}
\def\section{\@startsection{section}{1}{\z@}{3.5ex plus 1ex minus .2ex}
 {2.3ex plus .2ex}{\large\bf}}
\def\subsection{\@startsection{subsection}{2}{\z@}{2.3ex plus .2ex}
 {2.3ex plus .2ex}{\bf}}

    \renewcommand{\baselinestretch}{1.4}
\def\l{\langle}
\def\r{\rangle}
\def\vsq#1{\vert\l{#1}\r\vert^2}
\def\p23{\vsq{{\bar\Phi}_{23}}}
\def\v32{\vsq{V_3}}
\def\h18{\vsq{H_{18}}}
\def\anomaly{{{g^2}\over{16\pi^2}}{1\over{2\alpha^\prime}}}
\begin{document}

\begin{titlepage}
\samepage{
\setcounter{page}{1}
\rightline{UFIFT-HEP-96-30}
\rightline{UMD-PP-97-99}
\rightline{\tt hep-ph/9703235}
\rightline{February 1997}
\vfill
\begin{center}
 {\Large \bf  Meeting the Constraint of \\Neutrino--Higgsino Mixing
     in Gravity Unified Theories\\}
\vfill
 {\large Alon E. Faraggi$^1$\footnote{
   E-mail address: faraggi@phys.ufl.edu}
   $\,$and$\,$ Jogesh C. Pati$^{2}$\footnote{
   E-mail address: pati@umdhep.umd.edu}\\}
\vspace{.12in}
 {\it  $^{1}$ Institute For Fundamental Theory, 
  University of Florida, 
  Gainesville, FL 32611 
\\}
\vspace{.05in}
 {\it  $^{2}$  Department of Physics,
	       University of Maryland,
	       College park, MD 20742 
\\}
\end{center}
\vfill
\begin{abstract}
  {\rm
In Gravity Unified Theories all operators that are consistent
with the local gauge and discrete symmetries
are expected to arise in the effective low--energy theory.
Given the absence of multiplets like 126 of $SO(10)$ in string models, and
assuming that $B-L$ is violated spontaneously to generate
light neutrino masses via a seesaw mechanism, it is
observed that string theory solutions generically
face the problem of producing an excessive $\nu_L-{\tilde H}$
mixing mass at the GUT scale, which is some nineteen orders of magnitude
larger than the experimental bound of 1 MeV.
The suppression of $\nu_L-{\tilde H}$ mixing, like proton
longevity, thus provides one of the most severe constraints
on the validity of any string theory solution.
We examine this problem in a class
of superstring derived models. We find a family of solutions
within this class for which the symmetries of the models
and an allowed pattern of VEVs, surprisingly, succeed in
adequately suppressing
the neutrino--Higgsino mixing terms. At the same time they
produce the terms required to generate small neutrino masses
via a seesaw mechanism.
}
\end{abstract}
\vfill
\smallskip}
\end{titlepage}

\setcounter{footnote}{0}

\def\beq{\begin{equation}}
\def\eeq{\end{equation}}
\def\beqn{\begin{eqnarray}}
\def\eeqn{\end{eqnarray}}
\def\AEF{A.E. Faraggi}
\def\NPB#1#2#3{{\it Nucl.\ Phys.}\/ {\bf B#1} (19#2) #3}
\def\PLB#1#2#3{{\it Phys.\ Lett.}\/ {\bf B#1} (19#2) #3}
\def\PRD#1#2#3{{\it Phys.\ Rev.}\/ {\bf D#1} (19#2) #3}
\def\PRL#1#2#3{{\it Phys.\ Rev.\ Lett.}\/ {\bf #1} (19#2) #3}
\def\PRT#1#2#3{{\it Phys.\ Rep.}\/ {\bf#1} (19#2) #3}
\def\MODA#1#2#3{{\it Mod.\ Phys.\ Lett.}\/ {\bf A#1} (19#2) #3}
\def\IJMP#1#2#3{{\it Int.\ J.\ Mod.\ Phys.}\/ {\bf A#1} (19#2) #3}
\def\nuvc#1#2#3{{\it Nuovo Cimento}\/ {\bf #1A} (#2) #3}
\def\etal{{\it et al,\/}\ }
\hyphenation{su-per-sym-met-ric non-su-per-sym-met-ric}
\hyphenation{space-time-super-sym-met-ric}
\hyphenation{mod-u-lar mod-u-lar--in-var-i-ant}


\setcounter{footnote}{0}

While supersymmetry appears to be a key 
ingredient for higher unification, 
it is known that it poses two generic 
problems pertaining
to non--conservation of baryon and lepton numbers.
First and the better discussed is the problem 
of rapid proton decay, which 
arise through $d=4$ and color--triplet mediated and/or gravity
induced $d=5$ operators \cite{weinberg}. A problem of a similar magnitude, 
which is however not so well emphasized in the literature
is the one of excessive neutrino--Higgsino mixing, which 
could arise effectively through bilinear operators and could 
induce unacceptably large masses for the neutrinos. 
Using standard notation the relevant $(B,L)$ violating 
operators, which may arise in the superpotential, 
while respecting the Standard Model gauge symmetries,
are as follows :
\begin{eqnarray}
W = (\xi M^\prime) L H_2 &+& \left[\eta_1 {\bar U}{\bar D}{\bar D}+
		       \eta_2 QL{\bar D}+ \eta_3LL{\bar E}\right]\nonumber\\
&+& \left[\lambda_1 QQQL+\lambda_2{\bar U}{\bar U}{\bar D}{\bar E}
+\lambda_3 LL{H_2}{H_2}\right]/M\label{eq1}
\end{eqnarray}
Here, generation, $SU(2)_L$ and $SU(3)_C$ indices are suppressed. 
$M$ and $M^\prime$ denote characteristic mass scales. 
Experimental lower limits on proton life--time 
($\tau_{\rm p}> 10^{32} {\rm yrs}$)
turn out to impose the constraints \cite{hinchliff}:
$\eta_1\eta_2\le10^{-24}$ and ${\lambda_{1,2}/M}\le10^{-25}{\rm GeV}^{-1}$. 
Now, Eq. (\ref{eq1}) would also give rise to a $\nu-{\tilde H}_2$
mixing mass $\delta m_{\nu_H}=\xi M^\prime$, which in conjunction
with a Majorana mass for ${\tilde H}_2$ of order
$v_{\rm wk}\sim 100 {\rm GeV}$, would induce Majorana masses
for $\nu_L^\prime$s of order $(\xi M^\prime)^2/v_{\rm wk}$
if $\xi M^\prime<<v_{\rm wk}$. Astroparticle
physics restrictions on masses of stable neutrinos show that 
$m_{\nu_L}\le 10 {\rm eV}$, which implies that 
$\delta{m_{\nu_H}}=\xi M^\prime\le1 {\rm MeV}$\footnote{imposing
$m(\nu_e)<10^{-3} {\rm eV}$, for MSW type solutions for the 
solar neutrinos puzzle, yields $\delta m_{\nu_H}=\xi M^\prime<10 {\rm KeV}$.}.

Supersymmetry thus confronts us with two major 
puzzles: ({\bf I}) Why is the magnitude of the $d=4$ and $d=5$
operators so extraordinarily small, {\it i.e.} why $\eta_1\eta_2\le10^{-24}$ 
as opposed to being ${\cal O}(1)$, and 
${\lambda_{1,2}/M}\le10^{-25}{\rm GeV}^{-1}$, as opposed to 
being ${\cal O}(1 {\rm TeV}^{-1})$ ? and ({\bf II}) 
why is the neutrino--Higgsino
mixing mass so small -- $<1 {\rm MeV}$ or even $10 {\rm KeV}$, 
as opposed to being given, for example, by the string scale
or the $(B-L)$ breaking scale which is expected to be superheavy
($\ge10^{12} {\rm GeV}$, see below)? 

We would like to stress that these problems are not technical
but rather fundamental. In the context of point--particle quantum field
theories one can impose discrete and global symmetries, which 
forbid the undesired couplings. However, these type of symmetries
are in general expected to be violated by quantum gravity effects, 
unless they are obtained from broken gauge symmetries. Furthermore, 
the need to impose such ad hoc symmetries clearly indicates a 
structure which goes beyond conventional Grand Unified Theories. 
Indeed, it has been observed that a class of string models, possessing
three families, not only provide a natural doublet--triplet splitting
mechanism \cite{ps}, but also possess the desired gauge symmetries,
beyond SUSY GUTS, which safeguard proton stability from all potential
dangers \cite{custodial,pati}, including those which may arise from higher
dimensional operators and the color triplets in the infinite tower
of string states. 

The purpose of this note is to probe into the second puzzle
and to explore whether the strong suppression of $\nu_L-{\tilde H}$
can be understood in the context of string derived models 
in a manner similar to the understanding of proton stability. 

To appreciate the nature of the problem, we note that 
the $LH_2$--term, as also the $d=4$ operator in Eq. (\ref{eq1})
(though not the $d=5$), would of course be forbidden if 
one imposes a multiplicative R--parity symmetry: ${\rm R}=(-1)^{3(B-L)}$, 
which would have a natural origin through a gauged $(B-L)$ symmetry, 
as in $SO(10)$. The difficulty is that $SO(10)$ is expected
to be violated spontaneously at some heavy 
intermediate scale, especially if the neutrinos acquire light
masses through a seesaw mechanism, which 
assigns heavy Majorana masses to $\nu_R^\prime$s -- perhaps 
of order $10^{12}$GeV. Once $(B-L)$ is violated 
by the VEV of a scalar field, effective $LH_2$ term
and $d=4$ proton decay operators can in general be induced
through higher dimensional operators ($d\ge4$ and $d\ge5$ respectively)
by utilizing the VEV of such a field. Typically, 
the strength of such induced terms would be too large
(see below), unless symmetries beyond $(B-L)$
provide the desired protection. That such 
additional symmetries arise naturally in the 
context of a desirable class of three generation 
superstring models which suppress adequately the 
$d=4$ and $d=5$ proton decay operators was noted in Ref. \cite{pati}.
Here we examine whether string symmetries, beyond $(B-L)$
can adequately suppress the effective $LH_2$--term. 

It is worth noting at this stage that if $(B-L)$ is violated
by the VEV of the 126 of $SO(10)$ (or equivalently (10,3,1)
of $SU(4)\times SU(2)_L\times SU(2)_R$), then effectively 
R--parity=$(-1)^{3(B-L)}$ would still be preserved because 
this VEV violate $(B-L)$ by two units \cite{rnm}. in this case, 
the effective $LH_2$--term cannot be induced even if $(B-L)$
is broken. Recent works show, however, that 126 and 
very likely (10,3,1) as well, are hard -- perhaps impossible --
to obtain in string theories \cite{dmr}.
We will proceed by assuming that this constraint holds. 

In the absence of 126 of Higgs, $(B-L)$ can still be broken
and $\nu_R^\prime$s can acquire heavy Majorana masses, quite
simply by utilizing the VEVs of the sneutrino--like fields
${\tilde{\bar N}}$ and and ${\tilde N}^\prime_L$, which belong 
to the 
$16_H$ and ${\overline{16}}_H$ of $SO(10)$ respectively. 
The subscript ``$H$'' signifies that the corresponding $16$ 
is Higgs--like, which in general need not, and very likely does not,
coincide with the familiar chiral $16^\prime$s. Such 
vector--like representations (i.e. pairs of 16 and $\overline{16}$, 
some of which may not be Higgs--like), together with unpaired
chiral $16^\prime_i$s, do in fact arise generically in a large class 
of string derived models (see {\it e.g.} Ref. \cite{revamp,alr,tye}).
In this case, an effective 
operator of the form 
$16\cdot16\cdot{\overline{16}}_H{\overline{16}}_H/M$, 
which is allowed by $SO(10)$, would induce a Majorana mass
$({\bar\nu}_RC^{-1}{\bar\nu}_R^T)(\langle{\tilde N}^\prime_L\rangle
\langle{\tilde N}^\prime_L)/M+h.c.$ of magnitude $M_R\sim10^{12.5}$
GeV, as desired, for $\langle{\tilde N}^\prime_L\rangle\sim10^{15.5}$
GeV and $M=M_{\rm st}\sim 10^{18}$ GeV. With the sneutrino 
acquiring a VEV however, there is a danger that an effective
$d=4$ operator of the form $16_i16_H10_H$, which is allowed
by $SO(10)$ can lead to neutrino--Higgsino (i.e. $\nu_L-{\tilde H}_2$)
mixing through the coupling 
\begin{equation}
LH_2\langle{\tilde{\bar N}}\rangle
\label{nhmix}
\end{equation}
which is unacceptably large, if 
$\langle{\tilde{\bar N}}\rangle\sim 10^{15.5}$ GeV. 

As mentioned before
in the context of point particle field theory models, the problem 
can be resolved simply by imposing 
a discrete symmetry which forbids the Yukawa couplings of 
$16_H$ to the observed chiral families $(16)_i$ (this is in fact 
the reason why $16_H$ should not be identified with 
any of the chiral families, since the later must 
have large enough Yukawa couplings with each other to generate 
observed fermion masses and mixings). 

In any theory linked with gravity, like superstring theory, however, 
one can not impose such a discrete symmetry by hand. Rather, 
one must examine whether such a symmetry emerges from within the 
underlying theory. Furthermore, even if the cubic level
Yukawa coupling $16_i16_H10_H$ is absent, it may in general
be induced through gravity induced higher dimensional operators, 
by utilizing VEVs of relevant fields. One must thus examine 
whether intrinsic symmetries of the underlying theory -- e.g. 
a superstring theory -- together with an allowed pattern
of VEVs of the massless fields, which emerge from such a 
theory, can provide the needed enormous suppression
for $\nu_L-{\tilde H}$ mixing. {\it In this sense, 
the neutrino--Higgsino mixing, like proton stability, 
imposes a highly nontrivial constraint on all theories 
which incorporate gravity, especially if multiplets 
like 126 are not available to break $B-L$.}

In the following, we examine how the problem can be resolved in the 
context of some string derived solutions. First we note that for 
the free fermionic constructions yielding either the 
flipped $SU(5)\times U(1)$ \cite{revamp} or the $SO(6)\times SO(4)$ \cite{alr} 
models, there appear in addition to the three chiral families, 
additional massless vector--like pairs 
as components of $16$ and ${\overline{16}}$ (for $SU(5)\times U(1)$)
or half of $16$ and ${\overline{16}}$ (for $SO(6)\times SO(4)$)
containing additional right--handed neutrinos -- i.e.
${\bar N}_R$ and $N_L^\prime$, respectively. These additional 
vector--like pairs typically combine utilizing
VEVs of singlets to become superheavy ($\sim10^{15}-10^{15.5}$GeV)
while their sneutrino components ${\tilde{\bar N}}$
and ${\tilde N}^\prime$ acquire VEVs of the order of the GUT 
or string scales and thereby break the extended symmetries. 
As alluded to above, in these models excessive 
neutrino--Higgsino mixing will in general appear through
$16_i16_H10_H$ coupling\footnote{here we are using $SO(10)$
notation for convenience only. The corresponding coupling
in terms of the subgroups $SU(5)\times U(1)$
and $SO(6)\times SO(4)$ can be written unambiguously}
which may be either primary or induced via higher 
dimensional operators, unless suitable 
discrete, or continuous symmetries, forbid such couplings,
as well as the higher dimensional operators to a
sufficient degree. 
It is of great interest, and of crucial importance,
to examine whether the string derived 
models of the type mentioned above,
for which the sneutrino--like fields
must acquire large VEVs, posses the desired symmetries
to sufficiently suppress the $\nu_L-{\tilde H}$ mixing. 

We next turn our attention to the
class of superstring--derived standard--like models, obtained 
in ref. \cite{slm,eu}. 
A priori, motivations for exploring
this class of solutions are that -- (a) they exhibit qualitatively
the right texture for fermion masses and mixing, and
(b) they possess extra symmetries,
beyond conventional GUTs, which safeguard proton stability
from all potential dangers. These symmetries also turn out to be
helpful in suppressing $\nu_L-{\tilde H}$ mixing operators.
Furthermore, unlike the models of Refs. \cite{revamp,alr},
the standard--like models possess fields carrying half--integer
$B-L$ (see table 3).
Thus, they may well break $B-L$ by utilizing VEVs
of such fields rather than those of sneutrino--like fields.
In this case, $\nu_L-{\tilde H}$ mixing can still occur,
but only by utilizing products of VEVs
of such fields and of composites. 
As a result, the mixing necessarily involves higher 
dimensional operators, which are suppressed. However,
it is still far from clear as to whether $\nu_L-{\tilde H}$
mixing can be suppressed adequately in these models. This is what we
examine next. 

First we need to recall some salient features of the standard--like models.
This class of models is constructed by using the four dimensional
free fermionic construction \cite{FFF}. The standard--like models
are constructed by a set of eight boundary condition 
basis vectors $\{{\bf 1},S,b_1,b_2,b_3,\alpha,\beta,\gamma\}$ \cite{slm}.
The gauge symmetry of this class of models at the string
scale, after application of all GSO projections, has the form :
\begin{equation}
{\cal G}=\left[SU(3)_C\times SU(2)_L\times U(1)_{B-L}\times U(1)_{T_{3_R}}
         \right]\times\left[G_M={\prod_{i=1}^6}U(1)_i\right]\times
	 G_H\label{slmgroup}
\end{equation}
The first bracket denotes the part of the observable gauge symmetry
which is a subgroup of $SO(10)$. The $U(1)_i$ denote six horizontal
symmetry charges which act non--trivially on the three chiral families
as well as the hidden matter states. In the models of Ref. \cite{slm}, 
$G_H=SU(5)_H\times SU(3)_H\times U(1)_H^2$, which denotes 
the gauge symmetry of the hidden sector. 

Below we examine in detail the model of Ref. \cite{eu}.
The full massless spectrum of this model is given in
Ref. \cite{eu}. A partial list which is relevant for our 
purposes is given in tables 1--3. It includes the following
states : 

(I) There are three chiral families of quarks 
and leptons, each with sixteen components, including ${\bar\nu}_R$, 
which arise from the twisted sectors $b_1$, $b_2$ and $b_3$. 
These transform as 16's of $SO(10)$ and are neutral under $G_H$. 

(II) the Neveu--Schwarz (NS) sector corresponding to the
untwisted sector of the orbifold model produces, in addition to the 
gravity multiplets, three pairs of electroweak scalar 
doublets $\{h_1,h_2,h_3,{\bar h}_1,{\bar h}_2,{\bar h}_3\}$,
three pairs of $SO(10)$ -- singlets with $U(1)_i$ charge
$\{\Phi_{12},\Phi_{23},\Phi_{13},{\bar\Phi}_{12},{\bar\Phi}_{23},
{\bar\Phi}_{13}\}$, and three scalars that are singlets 
of the entire four dimensional gauge group, $\{\xi_1,\xi_2,\xi_3\}$.

(III) the sector $S+b_1+b_2+\alpha+\beta$ produces one additional pair
of electroweak doublets $\{h_{45},{\bar h}_{45}\}$, one pair of color
triplets $\{D_{45}, {\bar D}_{45}\}$ and seven pairs of $SO(10)$
singlets with $U(1)_i$ charge $\{\Phi_{45},{\bar\Phi}_{45},
\Phi_{1,2,3}^\pm,{\bar\Phi}_{1,2,3}^\pm\}$. 

(IV) In addition to the states mentioned above, which transform solely
under the observable gauge group and $U(1)_i$, the sectors $b_j+2\gamma+
(I={\bf 1}+b_1+b_2+b_3)$ produce hidden--sector multiplets
$\{T_i,{\bar T}_i,V_i,{\bar V}_i\}_{i=1,,2,3}$ which are $SO(10)$
singlets but are non--neutral under $U(1)_i$ and the hidden $G_H$.
The $T_i({\bar T}_i)$ are 5$(\bar5)$ and $V_i({\bar V}_i)$
are 3$(\bar3)$ of $SU(5)_H$ and $SU(3)_H$ gauge groups, respectively.
These states are listed in table 2.

(V) The vectors in some combinations of
$(b_1,b_2,b_3,\alpha,\beta)\pm\gamma+(I)$ produce additional states
which are either singlets of $SU(3)\times SU(2)\times U(1)_Y
\times SU(5)_H\times SU(3)_H$
or in vector--like representation of this group. The relevant states
of this class $\{H_{17}-H_{26}\}$ are listed in table 3. 
As we will show below the states of class (V) are crucial for
the resolution of the neutrino--Higgsino problem in the
superstring standard--like models. 

One characteristic feature of this class of models, 
is that, barring the three chiral 16's
there are no additional vector--like 
$16+{\overline{16}}$ pairs. As a result, elementary fields with the
quantum numbers of $N_L'\in{\overline{16}}$ do not exist
in this class of models.
Nevertheless, VEVs of products of certain condensates, 
which are expected to form through the hidden sector
force and certain fields belonging to the set (V)
can provide the desired quantum numbers of sneutrino
like fields -- i.e ${\bar N}_R\in 16$ and ${N_L'}\in{\overline{16}}$, 
as, for example, in the combinations shown below : 
\begin{eqnarray}
&&\langle H_{19}{\bar T}_i\rangle\langle H_{23}\rangle\rightarrow 
(B-L=-1,T_{3_R}=1/2)\sim N_L'\in {\overline{16}}\nonumber\\
&&\langle H_{20}{T}_i\rangle\langle H_{26}\rangle\rightarrow 
(B-L=+1,T_{3_R}=-1/2)\sim {\bar N}_R\in {{16}}
\label{effectiveN}
\end{eqnarray}
Note $H_{19}$ and $T_i$ transform as ${5}$, and $H_{20}$ and 
$T_i$ transform as 5, of $SU(5)_H$, respectively. 
Thus, $H_{19}(H_{20})$ can pair up with
${\bar T}_i(T_i)$ to make condensates 
at the scale $\Lambda_H$, where $SU(5)_H$ force becomes strong. 
In this model an effective 
seesaw mechanism \cite{seesaw,indirectseesaw}
is implemented 
by combining the familiar Dirac masses of the neutrinos which arise
through electroweak--symmetry breaking
scale, with superheavy mass terms which mix
${\bar \nu}_R^i$ with the singlet $\phi$ fields in sets
(II) and (III) \cite{fh}. The details of how the seesaw
mechanism arises in the superstring model are given in
Ref.\ \cite{fh}. Here we briefly sketch the 
main features. Subject to the relevant symmetries, the mixing
terms arise in order $N=6$, 7 and 8 for ${\bar\nu}_{2_R}$, 
${\bar\nu}_{3_R}$ and ${\bar\nu}_{1_R}$ respectively by utilizing VEVs 
of fields having the quantum numbers of $N_L'$ (see Eq. (\ref{effectiveN})). 
For example, for $N=6$, the mixing arises through the operators
${\bar\nu}_{2_R}{\Phi}_2^+\langle H_{19}{\bar T}_2\rangle 
\langle H_{23}\rangle
\langle\Phi_{45}\rangle$ while for $N=7$ and 8, the relevant terms
are given by ${\bar\nu}_{3_R}{\bar\Phi}_3^-
\langle H_{19}{\bar T}_3\rangle{\langle H_{25}\rangle}
\langle\Phi_{45}\rangle\langle{\bar\Phi}_{13}\rangle$ and
${\bar\nu}_{1_R}\xi_2\langle H_{19}{\bar T}_1\rangle
{\langle H_{25}\rangle}\langle\Phi_{45}\rangle\langle
{\bar\Phi}_{13}\rangle$. 
Thus, the neutrino mass matrix for a given family
takes the form 
\begin{equation}
{\left(\matrix{
                 {\nu_i}&{\nu_j^C}&{\phi_m}
                }
   \right)}
  {\left(\matrix{
                 0&(km_{_U})_{ij}&0\cr
                 (km_{_U})_{ij}&0&M_{\chi}\cr
                 0&M_{\chi}&O(M_\phi)\cr
                }
   \right)}
  {\left(\matrix{
                 {\nu_i}  \cr
                 {\nu_j^C}\cr
                 {\phi_m} \cr
                }
   \right)},
\label{nmm}
\end{equation}
with $m_\chi \sim  \left( \Lambda_{H}\over M \right)^2\left({\langle
\phi \rangle }\over M \right)^n M$ and $M_\phi \sim \left(\Lambda_{H}
\over M \right)^4 \left(\langle \phi \rangle \over M \right)^m M$.
$n$ and $m$ are the orders at which the terms for a given
neutrino flavor are obtained.  
The mass eigenstates are mainly $\nu$, $\nu_j^c$
and $\phi$ with a small mixing and with the eigenvalues
$$m_{\nu_j} \sim m_\phi \left({{k m_u} \over m_{\chi}}\right)^2
\qquad m_{\nu^c_j},m_{\phi} \sim m_{\chi}$$

Given that there are VEVs with the quantum numbers of $N_L'$
and ${\bar N}_R$ belonging to the ${\overline{16}}$ and 16
respectively (see Eq. (\ref{effectiveN})), the latter can lead to the 
$\nu_L{\tilde H}$ mixing through
higher dimensional operators. In the model of ref. \cite{eu}
we find that, subject to the constraints of the string--derived
symmetries and the string selection rules \cite{kln}, 
the relevant terms arise only at $N=6$, which we list
below : 
\begin{eqnarray}
L_1{\bar h}_1H_{26}H_{20}{T}_1\Phi_{13}~,\nonumber\\
L_1{\bar h}_2H_{24}H_{20}{T}_1\xi_1~,\nonumber\\
L_1{\bar h}_{45}H_{26}H_{20}{T}_1\Phi_{45}~,\nonumber\\
L_2{\bar h}_1H_{26}H_{20}{T}_2\Phi_{13}~,\nonumber\\
L_2{\bar h}_2H_{24}H_{20}{T}_2\xi_1~,\nonumber\\
L_2{\bar h}_2H_{18}H_{20}{T}_2\Phi_{45}~,\nonumber\\
L_2{\bar h}_{45}H_{26}H_{20}{T}_2\Phi_{45}~,\nonumber\\
L_3{\bar h}_{45}H_{17}H_{17}{V}_3{\bar\Phi}_{1}^-~,\nonumber\\
L_3{\bar h}_{2}H_{24}H_{20}{T}_3\xi_2~,\nonumber\\
L_3{\bar h}_{2}H_{18}H_{20}{T}_3\Phi_{45}.\label{n6terms}
\end{eqnarray}
It may be verified using tables 1--3 that all the terms
listed above conserve the full gauge symmetry listed in
Eq. (\ref{slmgroup}), including all the $U(1)'s$. The
gauge singlet fields $\xi_{1,2}$ appear in these terms to
produce non--vanishing Ising model correlators \cite{kln}. 

The superstring model under consideration contains 
an anomalous $U(1)$ symmetry which induces a Fayet--Iliopoulos
D--term and destabilizes the vacuum \cite{dsw}.
To preserve supersymmetry at the Planck scale, 
one must satisfy the $D$ and $F$ constraints 
arising from the superpotential by giving VEVs to a 
some scalar fields in the massless string spectrum. 
Depending on the choice of these VEVs at the 
string scale, the dangerous terms may indeed be generated. 
Thus, our task is to find a solution to the $F$ and $D$
constraints for which the neutrino--Higgsino mixing terms 
vanish while the seesaw neutrino mass terms, as well as other 
desirable phenomenological properties are retained.
Note that owing to differing quantum numbers (e.g. $B-L$ and $T_{3_R}$)
of ${\bar\nu}_R$ and $\nu_L$, the fields needed for
${\bar\nu}_R\phi$--mixing are distinct from those
which would induce $\nu_L-{\tilde H}$ mixing. For example, the former needs 
VEVs like those of $\{H_{23},H_{25},\Phi_{45},{\bar\Phi}_{13},
{\Phi}_1^-,\xi_2\}$ and the condensate $\l{H_{19}{\bar T}_i}\r$
while the latter needs VEVs like those of $\{H_{24},H_{26},H_{18},
\Phi_{45},\Phi_{13},\xi_1\}$ and the condensate $\l{H_{20}T_i}\r$.
Thus, if a suitable subset of the latter VEVs were zero,
while the former are non--zero, one could avoid $\nu_L-{\tilde H}$
mixing (at least up to N=6 terms), while allowing for the desired
${\bar\nu}_R^i\phi$--mixings. The catch is that one must, of course,
ensure that he desired pattern of VEVs is consistent with the $F$
and $D$--flat conditions, which is highly non--trivial.
This is what we proceed to do in the following.

The cubic level superpotential and the anomalous as
well as the anomaly free,
$U(1)$ combinations are given in ref. \cite{eu}. 
As an example, we find a solution to the 
$F$ and $D$ cubic level flatness constraints
with the following set of fields
\begin{equation}
\{
{\bar V}_2, {V}_3,
H_{18}, H_{23}, H_{25},
\Phi_{45}, {\bar\Phi}_1^-,\Phi_2^+,{\bar\Phi}_3^-,
{\bar\Phi}_{23}, {\bar\Phi}_{13}, \xi_1\},
\label{firstsolution}
\end{equation}
having non--zero VEVs and all other fields have vanishing VEV. 
With this set 
of fields the general solution is 
\begin{eqnarray}
&&\vsq{H_{23}}~=~\h18-\p23-{1\over6}\v32\label{h23}\\
&&\vsq{H_{25}}~=~\p23+{1\over6}\v32\label{h25}\\
&&\vsq{\Phi_{45}}~=~3\anomaly+\h18-{1\over{10}}\v32\label{p45}\\
&&\vsq{{\bar\Phi_{13}}}~=~\anomaly-{1\over5}\v32\label{p13}\\
&&\vsq{\Phi^+_2}~=~\anomaly-{8\over{15}}\v32\label{p2plus}\\
&&\vsq{{\bar\Phi}_3^-}~=~\anomaly-{1\over{30}}\v32\label{p3minus}\\
&&\vsq{{\bar\Phi}_1^-}~=~\anomaly-{8\over{15}}\v32\label{p1minus}\\
&&\vsq{{\bar V}_2}~=~\v32\label{v2}\\
&&\l{\xi_1}\r~=~-{{\l{\bar\Phi}_{23}\r\l{H_{25}}\r}
			\over{\l{H_{23}}\r}}\label{xsi}
\end{eqnarray}

In this solution the VEVs of three fields, $\{ V_3,
{\bar\Phi}_{23},H_{18}\}$ remain as
free parameters, which are restricted to give a positive
definite solution for the set of $D$--term equations.
Fixing those VEVs the above solution gives a qualitatively 
realistic fermion mass texture \cite{ckm}. 
As noted above, the model yields altogether four pairs
of electroweak Higgs--like doublets $\{h_1,h_2,h_3,h_{45}\}$
and $\{{\bar h}_1,{\bar h}_2,{\bar h}_3,{\bar h}_{45}\}$
(see sets {\bf{II}} and {\bf{III}}). It has been shown \cite{nrt}
that only one pair -- i.e. ${\bar h}_1$ or ${\bar h}_2$ and $h_{45}$, 
can remain light, while others acquire superheavy or intermediate
masses.  
Assuming that the light
electroweak doublets consist of ${\bar h}_1$ and
$h_{45}$ we observe that with this solution all the
neutrino--Higgsino mixing terms in Eq. (\ref{n6terms}) vanish while the 
seesaw terms in the neutrino mass matrix are preserved.
We checked with the aid of a computer program that with
this solution for the pattern of VEVs (Eq. \ref{firstsolution}),
surprisingly, the $\nu_L-{\tilde H}$ mixing terms are not induced
even up to $N=14$. One is led to suspect that very likely a 
discrete symmetry is left unbroken which prevents
$\nu_L-{\tilde H}$ mixing to all orders. Even if the mixing is
induced at $N=15$, however, we see that it would lead to a mixing mass
$\delta m_{\nu_H}\sim(\Lambda_H/M_{\rm st})^2
(\l\phi\r/M_{\rm st})^{10}\l\phi\r$, where $\Lambda_H$ denotes
the confinement scale of the hidden $SU(5)_H$ and $\l\phi\r$ is 
a weighted mean of the VEVs of 
fields listed in Eqs. (\ref{h23}--\ref{xsi}). For plausible values of 
$\l\phi\r/M_{\rm st}\sim(1/10-1/20)$, $\Lambda_H\sim10^{12}-10^{13}$GeV
and $M_{\rm st}\sim10^{18}$GeV we see that 
$\delta m_{\nu_H}\sim (10^{-10}-10^{-12})(10^{-13})(5\times10^{16}
{\rm GeV})\sim(1/2{\rm KeV}~{\rm to}1/2{\rm MeV})$.
Thus we see that the
symmetries of the string theory, together with an allowed pattern 
of VEVs, are powerful enough to provide a suppression of
$\delta m_{\nu_H}$ by as much as even 23 orders. This sort
of induced $\delta m_{\nu_H}$ is of course perfectly
compatible with the observational
limits of $\delta m_{\nu_H}<1{\rm MeV}$ (or even 10KeV). 
As noted above, while $\delta m_{\nu_H}$ is so strongly suppressed,
${\bar\nu}^i_R\phi$ -- mixing terms 
are still allowed and are given by $N=6$ ,7
and 8 terms for $i=1$, 2 and 3, mentioned before. These lead
to acceptable masses for the light $\nu_L^i$s.
A rough estimate (using Eq. (\ref{nmm})) yields:
$m(\nu_{\tau_L})\sim10^{-5}{\rm eV}>> m(\nu_{\mu_L})\ge
m(\nu_{e_L})$. In this case, if $\nu_L-{\tilde H}$
mixing mass is of order $10{\rm KeV}- 1{\rm MeV}$,
its contribution to $\nu_L$ masses would dominate and could
yield observable masses for the light neutrinos.  
The solution exhibited above thus
provides an example that string models can indeed yield qualitatively
realistic phenomenology, while suppressing the dangerous
neutrino--Higgsino mixing terms. 

We would like to point out
the important function of the exotic states
of class (V) in the solution to the neutrino--Higgsino mixing
problem. These states arise from sectors which break the $SO(10)$ 
symmetry to $SU(3)\times SU(2)\times U(1)^2$ \cite{ccf}. While they carry 
the standard charges under the standard model gauge group (and indeed
are Standard Model singlets) they carry non--standard charges under
the $U(1)_{Z^\prime}$ symmetry which is embedded in $SO(10)$ and
is orthogonal to the Standard Model weak hypercharge. It is
precisely due to this fact, as well as due to
the charges of these states under
the extra $U(1)$ symmetries, exhibited in Eq. (\ref{slmgroup}),
which provides the discrete symmetry
that are needed to suppress the neutrino--Higgsino mixing terms. 

To illustrate this point further, we would like to
mention in passing an
alternative scenario. In the solution above, to obtain
$D_{U(1)_{Z^\prime}}=0$ we used the fields, $H_{23}$, $H_{25}$ 
and the field with the opposite $U(1)_{Z^\prime}$ charge, $H_{18}$.
The field $H_{18}$ is the one which, combined with $H_{20}$,
gives the $U(1)_{Z^\prime}$ charge of $N_L^c$ and may therefore induce
the dangerous neutrino--Higgsino mixing terms. Suppose
that instead of assigning a VEV to $H_{18}$ we give a VEV to
$H_{20}$, or $H_{13}$, which carry the same charge under $U(1)_{Z^\prime}$.
These fields transform under the hidden non--Abelian $SU(5)_H$ 
and $SU(3)_H$ gauge groups, respectively, and would break
either of those gauge groups to a subgroup. 
In order to form a term which is invariant under all the 
gauge symmetries, a nonrenormalizable neutrino--Higgsino
mixing term must contain fields with the
quantum numbers of $H_{18}$ and ${\bar T}$.
Therefore, if we impose that all the fields with
the quantum numbers similar to $H_{18}$ have vanishing VEVs 
then there is a residual local discrete symmetry which 
forbids the neutrino--Higgsino mixing terms to all orders
of non--renormalizable terms. The vanishing of the D--term
equations for $U(1)_{B-L}$ and $U(1)_{T_{3_R}}$
and all the other $U(1)$'s in the observable sector
can still be satisfied with this choice of VEVs. 
In our superstring model
we find, however, that in this case the D--term equation for one of
the hidden $U(1)$ gauge groups cannot vanish (see Eq (\ref{slmgroup})).
We note, though, that this in itself may
not be a disaster as the communication to the observable sector
may produce sufficiently small supersymmetry breaking. 
Also the remaining freedom in the solution
of the D--term equations, as exhibited in Eqs. (\ref{h23}-\ref{xsi})
may allow the VEV of $H_{20}$ to be suppressed relative to 
the other VEVs,
thus allowing the actual scale of SUSY breaking in the hidden
sector to be suppressed relative to $M_{\rm st}$. 
We leave the investigation of this possibility to future work.
This scenario illustrates the crucial role of the
exotic states with ``fractional'' $U(1)_{Z^\prime}$ charge 
in suppressing the dangerous neutrino--Higgsino terms. 

In this paper we examined the problem of neutrino--Higgsino
mixing in superstring derived models. 
We stress that such a mixing is expected
to arise in any theory that aims 
at unification of gravity with the gauge interactions.
This mixing must be suppressed by at least 19 orders of
magnitude, compared to the GUT--scale, to conform with
observations. Therefore, neutrino--Higgsino mixing, like proton stability,
provides one of the most stringent constraints on the validity of
any gravity unified theory.
Furthermore, it is seen that the inclusion of
the expected gravitational effects, via the inclusion
of the higher order nonrenormalizable terms, affects
in a crucial way the phenomenology of the unified models.
We illustrated, in a specific string model, that
string theory can indeed produce qualitatively realistic
solutions, while the dangerous neutrino--Higgsino mixing terms
are sufficiently suppressed. In the solutions that we examined
the exotic ``stringy'' states play a crucial role.
These exotic states, of a uniquely stringy origin,
may have additional profound phenomenological implications
which merit further investigation.


It is a pleasure to thank K.S. Babu for most valuable discussions.
This work was supported in part by 
DOE Grant No.\ DE-FG-0586ER40272 (AEF) and by 
NSF Grant No.\ PHY--9119745 (JCP).

\bibliographystyle{unsrt}

\begin{thebibliography}{99}
\bibitem{weinberg} S. Weinberg,\PRD{26}{82}{475};\\
  S. Sakai and T. Yanagida, \NPB{197}{82}{533}. 
\bibitem{hinchliff} I. Hinchliffe, T. Kaeding, \PRD{47}{93}{279}.
\bibitem{ps}     A.E. Faraggi, \NPB{428}{94}{111}.
\bibitem{custodial}     A.E. Faraggi, \PLB{339}{94}{223}.
\bibitem{pati} J.C. Pati, \PLB{388}{96}{532}.
\bibitem{rnm} R.N. Mohapatra, \PRD{34}{86}{3457};\\
		A. Font, L. Ibanez and F. Quevedo, \PLB{288}{89}{79}.
\bibitem{dmr} K. Dienes and J. March--Russell, \NPB{479}{96}{113}.
\bibitem{revamp} I. Antoniadis, J. Ellis, J. Hagelin, and D.V. Nanopoulos,
                  \PLB{231}{89}{65};\\
	J. Lopez, D.V. Nanopoulos, and K. Yuan, \NPB{399}{93}{654}.
\bibitem{alr} 	I. Antoniadis, G.K. Leontaris, and J. Rizos,
						\PLB{245}{90}{161};\\
		G.K. Leontaris, \PLB{372}{96}{212}.
\bibitem{tye} S. Chaudhuri, S.--W. Chung, G. Hockney, and J. Lykken, 
					\NPB{456}{95}{89};\\ 
		G. Aldazabal, A. Font, L.E. Ib\'a\~nez, and A. Uranga, 
					\NPB{452}{95}{3};\\
		Z. Kakushadze, S.H.H. Tye, \PRL{77}{96}{2612}; 
						       hep-th/9609027;
							hep-th/9610106.
\bibitem{slm} A.E. Faraggi, \NPB{387}{92}{239}.
\bibitem{FFF}  H. Kawai, D.C. Lewellen, and S.-H.H. Tye,
				\NPB{288}{87}{1};\\
		I. Antoniadis, C. Bachas, and C. Kounnas,
				\NPB{289}{87}{87}.
\bibitem{eu} A.E. Faraggi, \PLB{278}{92}{131}.
\bibitem{seesaw} M. Gell--Mann, P.Ramond and R. Slansky, in
{\it Supergravity}, ed. by D. Freedman and P. Van--Nieuwenhuizen
(North--Holland 1979), p. 315;\\
 T. Yanagida, Proc. Workshop
on ``Unified Theories and Baryon Number of the Universe'', eds.
Sawata and A. Sugamoto, KEK, Japan (1979); \\
R.N. Mohapatra and G. Senjanovic, \PRL{44}{80}{912}.
\bibitem{indirectseesaw} For indirect seesaw see {\it e.g.}
R.N. Mohapatra and J.W. valle, \PRD{34}{86}{1642}, and references therein.
\bibitem{fh} A.E. Faraggi and E. Halyo, \PLB{307}{93}{311}.
\bibitem{kln} S. Kalara, J.L. Lopez and D.V. Nanopoulos, 
\NPB{353}{91}{650}.
\bibitem{dsw} M. Dine, N. Seiberg and E. Witten, \NPB{289}{87}{585}.
\bibitem{ckm} A.E. Faraggi and E. Halyo, \NPB{416}{94}{63}.
\bibitem{ccf} S. Chang, C. Coriano and A.E. Faraggi, \NPB{477}{96}{65}.
\bibitem{nrt} \AEF, \NPB{403}{93}{101}; \NPB{407}{93}{57}.
\bibitem{ssm}
        A. Font, L.E. Ibanez, F. Quevedo and A. Sierra,
                                        \NPB{331}{90}{421};\\
        D. Bailin, A. Love and S. Thomas, \NPB{298}{88}{75};\\
        J.A. Casas, E.K. Katehou and C. Mu{\~n}oz, \NPB{317}{89}{171}.
\bibitem{FNY} A.E. Faraggi, D.V. Nanopoulos, and K. Yuan,
				\NPB{335}{90}{347}.
\bibitem{lykken} S. Chaudhoury, G. Hockney and J. Lykken, 
			\NPB{469}{96}{357}.			
\end{thebibliography}

\vfill
\eject

\textwidth=7.5in
\oddsidemargin=-18mm
\topmargin=-5mm
\renewcommand{\baselinestretch}{1.3}
\pagestyle{empty}
\begin{table}
\begin{eqnarray*}
\begin{tabular}{|c|c|c|rrrrrrrr|c|rr|}
\hline
  $F$ & SEC & $SU(3)_C\times SU(2)_L$&$Q_{C}$ & $Q_L$ & $Q_1$ & 
   $Q_2$ & $Q_3$ & $Q_{4}$ & $Q_{5}$ & $Q_6$ & 
   $SU(5)_H\times SU(3)_H$ & $Q_{7}$ & $Q_{8}$ \\
\hline
   $L_1$ & $b_1$      & $(1,2)$ & $-{3\over2}$ & $0$ &
   ${1\over2}$ & $0$ & $0$ & ${1\over 2}$ & $0$ & $0$ &
   $(1,1)$ & $0$ & $0$ \\
   $Q_1$ &            & $(3,2)$ & $ {1\over2}$ & $0$ &
   ${1\over2}$ & $0$ & $0$ & ${1\over 2}$ & $0$ & $0$ &
   $(1,1)$ & $0$ & $0$ \\
   $d_1$ &            & $({\bar 3},1)$ & $-{1\over2}$ & $-1$ &
   ${1\over2}$ & $0$ & $0$ & $-{1\over2}$ & $0$ & $0$ &
   $(1,1)$ & $0$ & $0$ \\
   ${N}_1$ &            & $(1,1)$ & ${3\over2}$ & $-1$ & 
   ${1\over 2}$ & $0$ & $0$ & $-{1\over 2}$ & $0$ & $0$ &
   $(1,1)$ & $0$ & $0$ \\
   $u_1$ &            & $({\bar 3},1)$ & $-{1\over2}$ & $1$ &
   ${1\over2}$ & $0$ & $0$ & ${1\over2}$ & $0$ & $0$ &
   $(1,1)$ & $0$ & $0$ \\
   ${e}_1$ &            & $(1,1)$ & ${3\over2}$ & $1$ & 
   ${1\over 2}$ & $0$ & $0$ & ${1\over 2}$ & $0$ & $0$ &
   $(1,1)$ & $0$ & $0$ \\
\hline
   $L_2$ & $b_2$      & $(1,2)$ & $-{3\over2}$ & $0$ &
   $0$ & ${1\over2}$ & $0$ & $0$ & ${1\over 2}$ & $0$ &
   $(1,1)$ & $0$ & $0$ \\
   $Q_2$ &            & $(3,2)$ & $ {1\over2}$ & $0$ &
   $0$ & ${1\over2}$ & $0$ & $0$ & ${1\over 2}$ & $0$ &
   $(1,1)$ & $0$ & $0$ \\
   $d_2$ &            & $({\bar 3},1)$ & $-{1\over2}$ & $-1$ &
   $0$ & ${1\over2}$ & $0$ & $0$ & $-{1\over2}$ & $0$ &
   $(1,1)$ & $0$ & $0$ \\
   ${N}_2$ &            & $(1,1)$ & ${3\over2}$ & $-1$ & 
   $0$ & ${1\over 2}$ & $0$ & $0$ & $-{1\over 2}$ & $0$ &
   $(1,1)$ & $0$ & $0$ \\
   $u_2$ &            & $({\bar 3},1)$ & $-{1\over2}$ & $1$ &
   $0$ & ${1\over2}$ & $0$ & $0$ & ${1\over2}$ & $0$ &
   $(1,1)$ & $0$ & $0$ \\
   ${e}_2$ &            & $(1,1)$ & ${3\over2}$ & $1$ & 
   $0$ & ${1\over 2}$ & $0$ & $0$ & ${1\over 2}$ & $0$ &
   $(1,1)$ & $0$ & $0$ \\
\hline
   $L_3$ & $b_3$      & $(1,2)$ & $-{3\over2}$ & $0$ &
   $0$ & $0$ & ${1\over2}$ & $0$ & $0$ & ${1\over 2}$ & 
   $(1,1)$ & $0$ & $0$ \\
   $Q_3$ &            & $(3,2)$ & $ {1\over2}$ & $0$ &
   $0$ & $0$ & ${1\over2}$ & $0$ & $0$ & ${1\over 2}$ & 
   $(1,1)$ & $0$ & $0$ \\
   $d_3$ &            & $({\bar 3},1)$ & $-{1\over2}$ & $-1$ &
   $0$ & $0$ & ${1\over2}$ & $0$ & $0$ & $-{1\over2}$ & 
   $(1,1)$ & $0$ & $0$ \\
   ${N}_3$ &            & $(1,1)$ & ${3\over2}$ & $-1$ & 
   $0$ & $0$ & ${1\over 2}$ & $0$ & $0$ & $-{1\over 2}$ & 
   $(1,1)$ & $0$ & $0$ \\
   $u_3$ &            & $({\bar 3},1)$ & $-{1\over2}$ & $1$ &
   $0$ & $0$ & ${1\over2}$ & $0$ & $0$ & ${1\over2}$ & 
   $(1,1)$ & $0$ & $0$ \\
   ${e}_3$ &            & $(1,1)$ & ${3\over2}$ & $1$ & 
   $0$ & $0$ & ${1\over 2}$ & $0$ & $0$ & ${1\over 2}$ & 
   $(1,1)$ & $0$ & $0$ \\
\hline
   $h_1$ & ${\rm NS}$ & $(1,2)$ & $0$ & $-1$ &
   $1$ & $0$ & $0$ & $0$ & $0$ & $0$ &
   $(1,1)$ & $0$ & $0$ \\
   $h_2$ & 			 & $(1,2)$ & $0$ & $-1$ &
   $0$ & $1$ & $0$ & $0$ & $0$ & $0$ &
   $(1,1)$ & $0$ & $0$ \\
   $h_3$ & 			 & $(1,2)$ & $0$ & $-1$ &
   $0$ & $0$ & $1$ & $0$ & $0$ & $0$ &
   $(1,1)$ & $0$ & $0$ \\
   $\Phi_{12}$ & 		 & $(1,1)$ & $0$ & $0$ &
   $1$ & $-1$ & $0$ & $0$ & $0$ & $0$ &
   $(1,1)$ & $0$ & $0$ \\
   $\Phi_{13}$ & 		 & $(1,1)$ & $0$ & $0$ &
   $1$ & $0$ & $-1$ & $0$ & $0$ & $0$ &
   $(1,1)$ & $0$ & $0$ \\
   $\Phi_{23}$ & 		 & $(1,1)$ & $0$ & $0$ &
   $0$ & $1$ & $-1$ & $0$ & $0$ & $0$ &
   $(1,1)$ & $0$ & $0$ \\
\hline
   $h_{45}$ & $b_1+b_2+$ & $(1,2)$ & $0$ & $-1$ &
   $-{1\over2}$ & $-{1\over2}$ & $0$ & $0$ & $0$ & $0$ &
   $(1,1)$ & $0$ & $0$ \\
   $D_{45}$ & $\alpha+\beta$ & $(3,1)$ & $-1$ & $0$ &
   $-{1\over2}$ & $-{1\over2}$ & $0$ & $0$ & $0$ & $0$ &
   $(1,1)$ & $0$ & $0$ \\
   $\Phi_{45}$ & 		& $(1,1)$ & $0$ & $0$ &
   $-{1\over2}$ & $-{1\over2}$ & $-1$ & $0$ & $0$ & $0$ &
   $(1,1)$ & $0$ & $0$ \\
   $\Phi_{1}^\pm$ & 		 & $(1,1)$ & $0$ & $0$ &
   ${1\over2}$ & $-{1\over2}$ & $0$ & $\pm1$ & $0$ & $0$ &
   $(1,1)$ & $0$ & $0$ \\
   $\Phi_{2}^\pm$ & 		 & $(1,1)$ & $0$ & $0$ &
   ${1\over2}$ & $-{1\over2}$ & $0$ & $0$ & $\pm1$ & $0$ &
   $(1,1)$ & $0$ & $0$ \\
   $\Phi_{3}^\pm$ & 		 & $(1,1)$ & $0$ & $0$ &
   ${1\over2}$ & $-{1\over2}$ & $0$ & $0$ & $0$ & $\pm1$ &
   $(1,1)$ & $0$ & $0$ \\
\hline
\end{tabular}
\label{matter1}
\end{eqnarray*}
\caption{Massless states which transform solely under the 
observable gauge group. 
$Q_C=3/2(B-L)$ and $Q_L=2T_{3_R}$.
In the NS and the $b_1+b_2+\alpha+\beta$
sectors contain also the conjugate states (${\bar h}_1$, etc.).
The NS sector contains additional three singlet states, $\xi_{1,2,3}$,
which are neutral under all the $U(1)$ symmetries.} 
\end{table}

\vfill
\eject

\begin{table}
\begin{eqnarray*}
\begin{tabular}{|c|c|c|rrrrrrrr|c|rr|}
\hline
  $F$ & SEC & $SU(3)_C\times SU(2)_L$&$Q_{C}$ & $Q_L$ & $Q_1$ & 
   $Q_2$ & $Q_3$ & $Q_{4}$ & $Q_{5}$ & $Q_6$ & 
   $SU(5)_H\times SU(3)_H$ & $Q_{7}$ & $Q_{8}$ \\
\hline
   $V_1$ & $b_1+2\gamma+$ & $(1,1)$ & $0$ & $0$ & $0$ & ${1\over 2}$ &
   ${1\over2}$ & ${1\over 2}$ & $0$ & $0$ & $(1,3)$ & $-{1\over 2}$ &
   ${5\over2}$ \\
   ${\bar V}_1$ & $(I)$ & $(1,1)$ & $0$ & $0$ & $0$ & ${1\over 2}$ & 
   ${1\over2}$ & ${1\over 2}$ & $0$ & $0$ & $(1,{\bar3})$ & ${1\over 2}$ &
   $-{5\over2}$ \\
   $T_1$ &                   & $(1,1)$ & $0$ & $0$ & $0$ & ${1\over2}$ &
   ${1\over2}$ & $-{1\over2}$ & $0$ & $0$ & $(5,1)$ & $-{1\over 2}$ &
   $-{3\over2}$ \\
   ${\bar T}_1$ &            & $(1,1)$ & $0$ & $0$ & $0$ & ${1\over 2}$ & 
   ${1\over 2}$ & $-{1\over 2}$ & $0$ & $0$ & $({\bar 5},1)$ & ${1\over 2}$ &
   ${3\over2}$ \\
\hline
   $V_2$ & $b_2+2\gamma+$ & $(1,1)$ & $0$ & $0$ & ${1\over 2}$ &
    $0$ & ${1\over 2}$ & $0$ & ${1\over 2}$ & $0$ & $(1,3)$ & $-{1\over 2}$ &
   ${5\over2}$ \\
   ${\bar V}_2$ & $(I)$      & $(1,1)$ & $0$ & $0$ & ${1\over 2}$ & 
    $0$ & ${1\over 2}$ & $0$ & ${1\over 2}$ & $0$ & $(1,{\bar3})$ & 
    ${1\over 2}$ &
    $-{5\over2}$ \\
   $T_2$ &  		     & $(1,1)$ & $0$ & $0$ & ${1\over 2}$ & 
    $0$ & ${1\over 2}$ & $0$ & $-{1\over2}$ & $0$ & $(5,1)$ & $-{1\over 2}$ &
   $-{3\over2}$ \\
   ${\bar T}_2$ &  	     & $(1,1)$ & $0$ & $0$ & ${1\over 2}$ & 
    $0$ & ${1\over 2}$ & $0$ & $-{1\over 2}$ & $0$ & $({\bar5},1)$ & 
    ${1\over 2}$ &
    ${3\over2}$ \\
\hline
   $V_3$ & $b_3+\gamma+$ & $(1,1)$ & $0$ & $0$ & ${1\over 2}$ &
   ${1\over2}$ & $0$ & $0$ & $0$ & ${1\over2}$ & $(1,3)$ & $-{1\over 2}$ &
   ${5\over2}$ \\
   ${\bar V}_3$ & $(I)$ & $(1,1)$ & $0$ & $0$ & ${1\over2}$ & 
   ${1\over2}$ & $0$ & $0$ & $0$ & ${1\over2}$ & $(1,{\bar3})$ & ${1\over 2}$ &
   $-{5\over2}$ \\
   $T_3$ &  		     & $(1,1)$ & $0$ & $0$ & ${1\over2}$ & 
   ${1\over2}$ & $0$ & $0$ & $0$ & $-{1\over 2}$ & $(5,1)$ & $-{1\over 2}$ &
   $-{3\over2}$ \\
   ${\bar T}_3$ &  	     & $(1,1)$ & $0$ & $0$ & ${1\over 2}$ & 
   ${1\over2}$ & $0$ & $0$ & $0$ & $-{1\over 2}$ & $({\bar5},1)$ & 
   ${1\over 2}$ &
   ${3\over2}$  \\
\hline
\end{tabular}
\label{matter2}
\end{eqnarray*}
\caption{Massless states from the sectors $b_j+2\gamma$. 
$Q_C=3/2(B-L)$ and $Q_L=2T_{3_R}$.} 
\end{table}

\vfill
\eject

\begin{table}
\begin{eqnarray*}
\begin{tabular}{|c|c|c|rrrrrrrr|c|rr|}
\hline
  $F$ & SEC & $SU(3)_C\times SU(2)_L$&$Q_{C}$ & $Q_L$ & $Q_1$ & 
   $Q_2$ & $Q_3$ & $Q_{4}$ & $Q_{5}$ & $Q_6$ & 
   $SU(5)_H\times SU(3)_H$ & $Q_{7}$ & $Q_{8}$ \\
\hline
   $H_{13}$ & $b_1+b_3+$ & $(1,1)$ & $-{3\over4}$ & ${1\over2}$ & 
   $-{1\over4}$ & ${1\over 4}$ & $-{1\over 4}$ & 0 & 0 & 0 & 
   (1,3) & ${3\over4}$ & $5\over4$   \\
   $H_{14}$ &  $\alpha\pm\gamma+$  & (1,1) & ${3\over4}$  & $-{1\over2}$ & 
   ${1\over4}$ & $-{1\over 4}$ & $1\over 4$ & 0 & 0 & 0 &
   (1,$\bar 3$) & $-{3\over 4}$ & $-{5\over 4}$  \\
   $H_{15}$ &     $(I)$       & (1,2) & $-{3\over4}$ & $-{1\over2}$ & 
   $-{1\over4}$ & ${1\over 4}$ & $-{1\over 4}$ & 0 & 0 & 0 &
   (1,1) & $-{1\over 4}$ & $-{15\over 4}$  \\
   $H_{16}$ &               & (1,2) & ${3\over4}$ & $1\over2$ & 
   ${1\over4}$ & $-{1\over 4}$ & $1\over 4$ & 0 & 0 & 0 &  
   (1,1) & $1\over4$ & ${15\over 4}$  \\
   $H_{17}$ &               &  (1,1) & $-{3\over4}$ & $1\over2$ &
   $-{1\over 4}$ & $-{3\over4}$ & $-{1\over 4}$ & 0 & 0 & 0 &
   (1,1) & $-{1\over 4}$ & $-{15\over 4}$  \\
   $H_{18}$ &               & (1,1) & $3\over4$ & $-{1\over2}$ &
   ${1\over 4}$ & $3\over4$ & $1\over 4$ & 0 & 0 & 0 &
   (1,1) & ${1\over 4}$ & ${15\over 4}$  \\
\hline
   $H_{19}$ & $b_2+b_3+$ & (1,1) & $-{3\over4}$ & $1\over2$ & 
   ${1\over 4}$ & $-{1\over4}$ & $-{1\over 4}$ & 0 & 0 & 0 &
   (5,1)&  $-{1\over 4}$ & ${9\over 4}$ \\
   $H_{20}$ & $\alpha\pm\gamma+$  & (1,1) & $3\over4$ & $-{1\over2}$ & 
   $-{1\over 4}$ & $1\over4$ & $1\over 4$ & 0 & 0 & 0 & 
   ($\bar 5$,1) & $1\over 4$ & $-{9\over 4}$  \\
   $H_{21}$ &  $(I)$         & (3,1) & $1\over4$ & $1\over2$ & 
   ${1\over 4}$ & $-{1\over 4}$ & $-{1\over4}$ & 0 & 0 & 0 & 
   (1,1) & $-{1\over 4}$ & $-{15\over 4}$  \\
   $H_{22}$ &                & ($\bar 3$,1) & $-{1\over4}$ & $-{1\over2}$ 
   & $-{1\over 4}$ & ${1\over 4}$ & $1\over4$ & 0 & 0 & 0 &
   (1,1) & ${1\over 4}$ & ${15\over 4}$  \\
   $H_{23}$ &                & (1,1) & $-{3\over4}$ & $1\over2$ & 
   ${1\over 4}$ & $-{1\over 4}$ & $3\over4$ & 0 & 0 & 0 &     
   (1,1) & ${1\over 4}$ & ${15\over 4}$  \\
   $H_{24}$ &                & (1,1) & $3\over4$ & $-{1\over2}$ & 
   $-{1\over4}$ & ${1\over 4}$ & $-{3\over4}$ & 0 & 0 & 0 &
   (1,1) & $-{1\over 4}$ & $-{15\over 4}$ \\
   $H_{25}$ &                & (1,1) & $-{3\over4}$ & $ 1\over2$ & 
   $1\over4$ & $3\over4$ & $-{1\over4}$ & 0 & 0 & 0 &
   (1,1) & $-{1\over4}$ & $-{15\over4}$ \\
   $H_{26}$ &                & (1,1) & $3\over4$ &  $-{1\over2}$ & 
   $-{1\over4}$ &  $-{3\over4}$ & $1\over4$ & 0 & 0 & 0  & 
   (1,1) & $1\over4$ & $15\over4$  \\
\hline
\end{tabular}
\label{matter3}
\end{eqnarray*}
\caption{The exotic massless states from the sectors
$b_1+b_3+\alpha\pm\gamma+(I)$ and $b_2+b_3+\alpha\pm\gamma+(I)$.} 
\end{table}

\end{document}